\documentclass[10pt,a4paper]{article}
\usepackage[latin1]{inputenc}
\usepackage{amsmath}
\usepackage{amsfonts}
\usepackage{amssymb}
\usepackage{graphicx}
\usepackage{upgreek}
\usepackage[affil-it]{authblk}

\author[1]{R. Thiru Senthil \thanks{rtsenthil@imsc.res.in}}
\author[2]{G. Rajasekaran \thanks{graj@imsc.res.in}}
\affil[1]{Homi Bhabha National Institute, Mumbai}
\affil[1,2]{The Institute of Mathematical Sciences, Chennai}
\affil[2]{Chennai Mathematical Institute, Chennai}

\title{Anomalous Kolar Events and Dark Matter Decay in Dwarf Spheroidal Galaxies}
\date{}
\begin{document}
\maketitle
\begin{abstract}
Using the Fermi LAT data on the gamma ray emission from dwarf spheroidal galaxies, we get the upper bound on the probability of gamma rays from dark matter decay for the validity of explanation of the anomalous Kolar events as dark matter decay.
\end{abstract}

\section{Introduction}

At the Kolar gold fields (KGF) in India, deep under ground particle physics experiments were carried out from 1960 to 1992.  During two separate periods of these experiments, some anomalous events were seen.  Five such events was reported during the cosmic ray neutrino experiments and three events during proton decay studies \cite{Krishnaswamy:1975zu} \cite{Krishnaswamy:1975qe} \cite{proceeding}.  In total, there were eight events which are not understandable with known physics.  These events are now known as anomalous Kolar events.

The possibility of explaining the anomalous Kolar events via decay of dark matter was recently pointed out \cite{Murthy:2013uca}.  Decay of neutral dark matter particle at rest in the mass range of $5$ to $10 ~GeV$ with life time of order around the age of the universe ($10^{17}s$) could explain these events.

If the local number density of dark matter particle (DMP) in the solar system is n, the decay rate of DMP is $ \Gamma $, the effective volume of the cavern is $ V $ and the branching ratio to the decay into visible channels is B, then the rate of events seen in the cavern is given by,
$$R = n \Gamma V B$$

If we take n to be the range of one per cc, $ \Gamma $ to be $ 10^{-17} s^{-1} $ (with DMP life time roughly around the age of the universe), V to be $ 10 m \times 10 m \times 10 m \approx 10^9 m^3$  and $ B \approx 1 $, we get $ R \approx 0.1 $ events per year.  All these numbers are very approximate.  Especially the volume V, since the cavern does not exist now.  It was remarkable that such a crude estimate agreed roughly with the rate of the anomalous events seen at Kolar.

If this explanation is correct, it would have solved two problems in one stroke - interpretation of the anomalous Kolar events and the observation of DMP.

But, after the above proposal was published a problem was noticed which arises from limits on the lifetime of decaying DMP due to the data from Fermi Large Area Telescope (LAT) gamma rays observation.  Based on the fact that no unambiguous signal for dark matter decay has been found in gamma ray observation, an estimate of the lower bound on the lifetime of DMP has been made in the range of $ \approx 10^{26} s $ for the mass range of $ 10$ to $10,000 ~GeV $.  Fermi LAT observations on dwarf spheroidal galaxies give tighter bounds by assuming certain particle physics models \cite{Baring:2015sza}.  These indirect astrophysical bounds appear much too high for the Kolar events to be interpreted as due to the decay of DMP through conventional channels.  In this note, we reexamine this problem.

\section{Dark matter decay in dwarf spheroidal galaxies}

Among various astrophysical sources, dwarf spheroidal galaxies (dSphs) are favourable for indirect detection of dark matter.  Although dSphs provides faint signal compared with other sources of indirect detection such as galactic cluster, dSphs have smaller astrophysics background for gamma ray observation.  They also have large mass to luminosity ratio ($\sim 1000 ~ order$) which shows that they have large dark matter content.  The deficiency for producing high energy photons in their intrinsic sources make dSphs as favourable clean environment for indirect dark matter detection.  

The Fermi LAT produced important results for indirect dark matter detection in verity of astrophysical sources including dSphs.  So far, there is no conclusive evidence for the observation of significant excess of gamma rays over known astrophysical background of dSphs.  But, these observations result in setting strongest constraints over dark matter properties for the assumed particle physics model.  For our analysis, we consider a set of 27 dSphs.  We obtain model independent constrains on particle physics model for dark matter decay from the result of gamma ray observations with Fermi LAT.

\section{Model independent decay constraints for dark matter in dwarf spheroidal galaxies}

  There are many literature available for constraining dark matter annihilation cross section and dark matter life time using Fermi LAT data.  These studies usually assume the production of gamma rays by considering dark matter interaction with themselves and with standard model particles.  The limits are obtained for different channels of consideration.  Model independent constraints on particle physics model for dark matter annihilation was calculated by Boddy et.al. \cite{Boddy:2018qur} from the analysis of around nine years of observation results of Fermi LAT and calculating background distribution for different sets of galaxies.  We obtain similar constraints on models for decaying dark matter using the Fermi LAT data and background distributions provided with their results.

\subsection{Analysis framework}
The number of photons expected from dark matter decay in dwarf spheroidal galaxy is,
\begin{equation}
N = \frac{\Gamma}{m_{\chi}} \times \Phi_{P} \times ~ J_{d} \times \left( A_{eff}T_{obs} \right).
\end{equation}
where,
\begin{description}
\item $\Gamma$ is the decay rate of dark matter
\item $m_{\chi}$ is mass of the dark matter particle
\item $A_{eff}$ is the effective area of the detector.  We ignore the energy dependence of effective area in the analysis.
\item $T_{obs}$ is the observation time.
\end{description}
and
\begin{equation}
\Phi_{P}  = \dfrac{1}{4 \pi } \int_{E_{th}}^{E_{max}} \sum_{f} B_{f} \dfrac{dN_f}{dE} dE
\end{equation}
Here,
\begin{description}
\item $\Phi_{P}$ is a factor proportional to the production of gamma rays in dark matter decay
\item $\dfrac{dN_{f}}{dE}$ = Gamma ray spectrum arising from the decay of dark matter for the decay channel $f$.
\item $B_{f}$  is the branching factor for the decay channel $f$.
\end{description}

\begin{description}
\item The total gamma ray photon flux emitted per decay is obtained from summing the spectral flux ($\dfrac{dN_{f}}{dE}$) over all possible final states and integrating in the energy range of $E_{th}$ threshold energy to $E_{max}$ maximum energy.
\end{description}

\begin{equation}
J_{d}= \int_{\Delta \Omega} \int \rho_{DM}(r,\Omega) dr d\Omega
\end{equation}
\begin{description}
\item $J_{d}$ contains information of dark matter distribution in dSph.
\item $r$ is the distance from the detector to the dSph under consideration along the line of sight.
\item  $\rho_{DM}$ is the dark matter density distribution within the region of interest ($\Delta \Omega$) which is usually calculated from spherically symmetric density distribution.
\end{description}

The factor ($\Phi_{P}$) is independent of astrophysics.  Several literature are available for the astrophysical factor ($J_d$) for decaying dark matter in different dSphs.  They are usually computed for an opening angle of $0.5^{0}$ by assuming Navarro Frenk White (NFW) profile for the dark matter distribution in dSph.  Fermi LAT observation data provides number of observed photons ($N_{obs}$) and ($A_{eff}T_{obs}$) for each dSph.  In stacking procedure, we consider a set of dSphs ($\left\lbrace dSphs \right\rbrace$).  We can then calculate the number of expected average number of photons ($N$) from decaying dark matter over astrophysical background distribution with confidence level ($\beta$) for this set of dSphs following the procedure by \cite{Boddy:2018qur}.  Since, $\Phi_{P}$ only depends on the model under consideration, the calculated bound on $N$ directly provides bound on $\Phi_{P}$ as given by equation \ref{eq:bound}.

\begin{equation}
\Phi_{P}\left( \beta \right) = \dfrac{m_{\chi} \times N \left( \beta \right)}{\Gamma \times \displaystyle\sum_{i \in \left\lbrace dSphs \right\rbrace} \left[ J_{d}^{i}(\Delta \Omega) \left(A_{eff}T_{obs}\right)^{i} \right] }
\label{eq:bound} 
\end{equation}

\subsection{Bound on models from dSph data} 

We consider a set of 27 dSphs for our analysis.  They are listed in table \ref{tab:galprop} along with their Fermi LAT data.  For this set of galaxies, the Fermi LAT observation $A_{eff}T_{obs}$, and Number of observed and background photons ($N_{obs}$ and $N_{bgd}$) and calculated empirical background distribution $\left( P_{bgd}^{i}\left( N_{bgd}^{i} \right) \right)$ are from \cite{Boddy:2018qur}.  These results were obtained from the analysis of Fermi LAT around 9 years of observation for the energy range 1 to 100 GeV and region of interest (ROI) of $0.5^0$ around each dSph.  The total number of background photons and observed photons in this set were $N_{bgd}^{Tot} = 3698$ and $N_{obs}^{Tot} = 3733$.

We adopt astrophysical factor $J_{d}$ for our set of dSphs from \cite{Evans:2016xwx}.  They are given in table \ref{tab:galprop}.  The stacking procedure to calculate the particle physics model bound is as follows:

\begin{itemize}
\item We calculate total background probability mass distribution function for our considered set of dSphs, from convolution of individual empirical background distributions of each dSph.  This provides distribution function for obtaining probability ($P_{bgd}$) for background photons ($N_{bgd}$).

\begin{equation}
P_{bgd}\left( N_{bgd} \right) = \displaystyle\sum_{\displaystyle\sum_{i}N_{bgd}^{i}=N_{bgd}} \prod_{i} P_{bgd}^{i}\left( N_{bgd}^{i} \right)
\end{equation}

The normalized total background probability distribution for our set is provided in figure \ref{fig:bgd27}.
\begin{figure}
	\includegraphics[width=\linewidth]{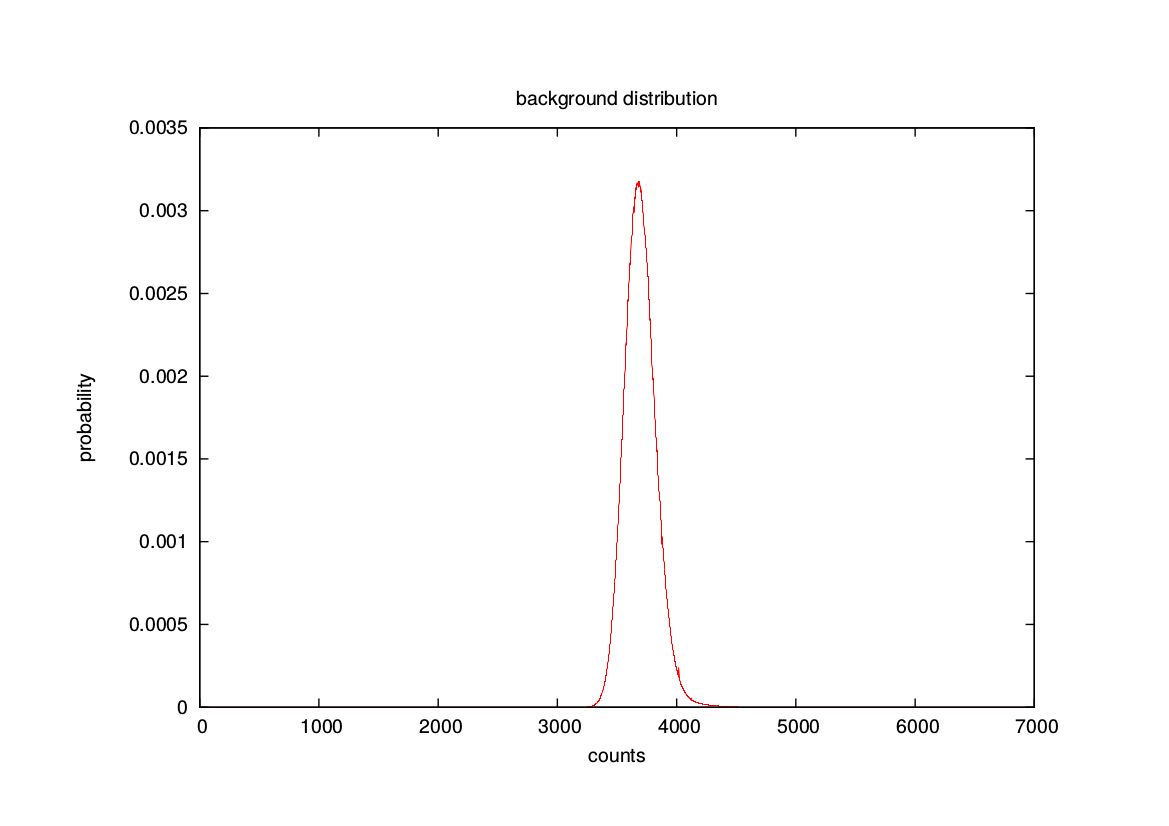}
	\caption{Total background distribution for our set of 27 dwarf spheroidal galaxies.}
	\label{fig:bgd27}
\end{figure}

\item We assume poisson distribution for number of photons expected from decaying dark matter.  We create distribution tables with mean value as expected number of photons which is $N_{DM}^{exp}$ .  If we assume a total of $N_{DM}^{exp}$ photons were coming from dark matter decay, we then have the probability distribution for $N_{DM}$ signals, from the corresponding signal distribution function $P \left( N_{DM} ; N_{DM}^{exp}\right)$.

\begin{equation}
P \left( N_{DM} ; N_{DM}^{exp}\right) = e^{- N_{DM}^{exp}} \dfrac{\left( N_{DM}^{exp}\right)^{N_{DM}}}{N_{DM}!}
\end{equation}

\item The convolution of signal and background provides the total distribution for gamma rays.  If we assume $N_{DM}^{exp}$ photons were originated from dark matter decay, then the probability for producing more than the $N_{obs}^{Tot}$ observed number of photons is,

\begin{equation}
\displaystyle\sum_{\left( N_{bgd}+N_{DM} \right) > N_{obs}^{Tot}} P_{bgd}\left( N_{bgd} \right) \times P \left( N_{DM} ; N_{DM}^{exp}\right)
\end{equation}

\item We can calculate upper bound on number of photons $N\left( \beta \right)$ originated from dark matter decay, with confidence level $\beta$ from,

\begin{equation}
\displaystyle\sum_{\left( N_{bgd}+N_{DM} \right) > N_{obs}^{Tot}} P_{bgd}\left( N_{bgd} \right) \times P \left( N_{DM} ; N \left( \beta \right)\right) = \beta 
\end{equation}

\end{itemize}

With $\beta$ confidence level, we therefore get the upper limit on $N_{DM}^{exp}$ as,
\begin{equation}
N_{DM}^{exp} < N \left( \beta \right)
\end{equation}

Since the knowledge of dark matter distribution in dSphs is limited, there are huge uncertainties in the calculation associated with $J_{d}$.  We treat them as systematic uncertainties.  This provide uncertainties in our calculation for $\Phi_{P} \left( \beta \right)$ for given confidence level $\beta$.

We choose $m_{\chi} = 10 ~GeV $ and $\Gamma \approx 10^{-17} s^{-1}$ since they are required for the explanation of the anomalous Kolar events.

\begin{figure}
	\includegraphics[width=\linewidth]{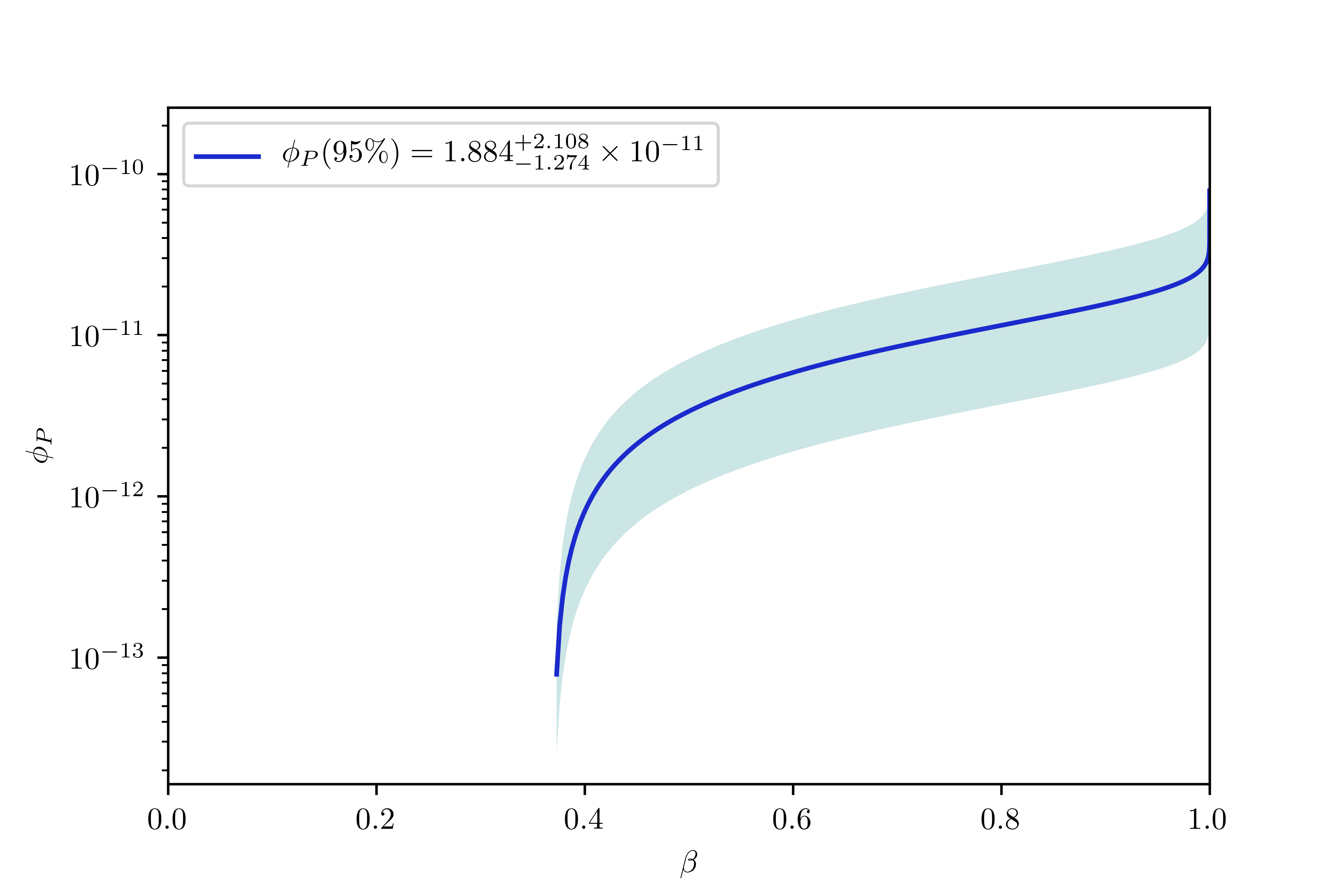}
	\caption{Upper bound on $\Phi_{P}$ for dark matter decay.  The solid line was obtained by considering the central value of decay factors $J_{d}$s for dSphs.  The edges of the band corresponds to $\pm 1 \sigma$ variations in the values of decay factors $J_{d}$}
	\label{fig:ppbound}
\end{figure}

By using equation \ref{eq:bound}, we calculate $\Phi_{P} \left( \beta \right)$ from $N \left( \beta \right)$ for chosen $m_{\chi}$ and $\Gamma$ with $1 \sigma$ systematic error provided with of $J_{d}^{i}$ of every dSph in our set.  The calculated $\Phi_{P} \left( \beta \right)$ for varying confidence level $\beta$ is given in figure \ref{fig:ppbound}.

For, $\beta = 0.95$ (95\%) confidence level, we obtained the bound on the factor $\phi_{P}$ as,
\begin{equation}
\phi_{P} \left( 95 \% \right)= 1.884 ^{+2.107}_{-1.274} \times 10^{-11}
\end{equation}

Figure \ref{fig:ppbound} gives the upper bound for the production of gamma rays in dark matter decay for various $\beta$, for the viability of the explanation of the anomalous Kolar events.  We have calculated the $\phi_{P}$ with dark matter mass $m_{\chi} = 10 ~GeV$.  We can obtain the upper bound $\phi_{P}$ for different dark matter mass $m_{\chi}$ (in GeV units) by multiplying with a scale factor $\frac{m_{\chi}}{10}$.


\begin{table}
\centering
\caption{Properties and Fermi LAT data of dwarf spheroidal galaxies considered in our analysis.  Decay factors $J_{d}$ within angular cone of $0.5^0$ for each dSph are from \cite{Evans:2016xwx}. Fermi LAT data analysis provide the average exposure $A_{eff}T_{obs}$, the number of gamma rays seen $N_{}obs$ and number of background photons $N_{bgd}$.  They were taken from \cite{Boddy:2018qur}. }
\vspace*{0.3cm}
\label{tab:galprop}
\begin{tabular}{|l|c|c|c|c|}
\hline
&&&&\\
dSph Name & $ log_{10} \left[ \frac{ J_{d} }{ GeV cm^{-2}} \right] $ & $A_{eff}T_{obs} $  $ cm^{2}s$ & $N_{bdg}$ & $N_{obs}$ \\
&&&&\\
\hline
 Bootes I	&$	17.28	^{+	0.64	}_{	-0.38	}	$&	4.042	* E11	&	137	&	128	\\
 Canes Venatici I	&$	17.78	^{+	0.11	}_{	-0.11	}	$&	4.27	* E11	&	102	&	72	\\
 Canes Venatici II	&$	17.37	^{+	0.4	}_{	-0.4	}	$&	4.259	* E11	&	103	&	91	\\
 Carina	&$	17.98	^{+	0.34	}_{	-0.34	}	$&	4.363	* E11	&	203	&	159	\\
 Coma Berenices	&$	18.06	^{+	0.32	}_{	-0.32	}	$&	4.046	* E11	&	115	&	151	\\
 Draco	&$	18.39	^{+	0.25	}_{	-0.25	}	$&	5.366	* E11	&	175	&	150	\\
 Fornax	&$	18.26	^{+	0.17	}_{	-0.17	}	$&	3.993	* E11	&	92	&	125	\\
 Grus I	&$	17.59	^{+	0.46	}_{	-0.96	}	$&	4.191	* E11	&	109	&	105	\\
 Hercules	&$	17.38	^{+	0.45	}_{	-0.45	}	$&	4.33	* E11	&	234	&	222	\\
 Horologium I	&$	17.78	^{+	0.47	}_{	-0.2	}	$&	4.394	* E11	&	110	&	132	\\
 Hydra II	&$	16.89	^{+	0.44	}_{	-0.92	}	$&	4.012	* E11	&	205	&	162	\\
 Leo I	&$	17.89	^{+	0.28	}_{	-0.28	}	$&	3.879	* E11	&	128	&	138	\\
 Leo II	&$	17.62	^{+	0.25	}_{	-0.25	}	$&	3.996	* E11	&	111	&	83	\\
 Leo IV	&$	17.22	^{+	0.9	}_{	-0.9	}	$&	3.67	* E11	&	131	&	133	\\
 Leo T	&$	17.35	^{+	0.37	}_{	-0.37	}	$&	3.993	* E11	&	130	&	122	\\
 Leo V	&$	17.23	^{+	1.05	}_{	-0.7	}	$&	3.682	* E11	&	130	&	145	\\
 Pisces II	&$	17.41	^{+	0.57	}_{	-0.4	}	$&	3.718	* E11	&	152	&	137	\\
 Reticulum II	&$	17.93	^{+	0.85	}_{	-0.32	}	$&	4.423	* E11	&	108	&	128	\\
 Sculptor	&$	18.33	^{+	0.29	}_{	-0.29	}	$&	3.897	* E11	&	88	&	114	\\
 Segue 1	&$	18.17	^{+	0.39	}_{	-0.39	}	$&	3.947	* E11	&	128	&	154	\\
 Segue 2	&$	17.08	^{+	0.86	}_{	-1.75	}	$&	4.072	* E11	&	210	&	246	\\
 Sextans	&$	18.07	^{+	0.29	}_{	-0.29	}	$&	3.699	* E11	&	131	&	139	\\
 Tucana II	&$	18.45	^{+	0.88	}_{	-0.58	}	$&	4.518	* E11	&	121	&	128	\\
 Ursa Major I	&$	18.15	^{+	0.25	}_{	-0.25	}	$&	4.823	* E11	&	110	&	108	\\
 Ursa Major II	&$	18.48	^{+	0.39	}_{	-0.39	}	$&	5.594	* E11	&	182	&	225	\\
 Ursa Minor	&$	18.45	^{+	0.24	}_{	-0.24	}	$&	5.701	* E11	&	146	&	123	\\
 Willman 1	&$	18.03	^{+	0.91	}_{	-0.62	}	$&	4.771	* E11	&	108	&	113	\\

\hline
\end{tabular}
\end{table}


\section{Summary}

Using the upper bounds on the number of gamma rays from the dSphs, we have obtained the upper bound on the probability of emission of gamma rays from dark matter decay.  The next step will be to construct model(s) for dark matter decay that will be consistent with this bound.  Such models would be the possible explanation for the observed anomalous Kolar events.

\section{Acknowledgements}
We acknowledge IMSc High Performance Computing (HPC) - Nandadevi cluster facility where our calculations were done.  We thank Tathagata Ghosh for discussions.



\begin{thebibliography}{1}

\bibitem{Krishnaswamy:1975zu} 
  M.~R.~Krishnaswamy, M.~G.~K.~Menon, V.~S.~Narasimham, N.~Ito, S.~Kawakami and S.~Miyake,
  Phys.\ Lett.\  {\bf 57B}, 105 (1975).
  doi:10.1016/0370-2693(75)90255-5

\bibitem{Krishnaswamy:1975qe} 
  M.~R.~Krishnaswamy, M.~G.~K.~Menon, V.~S.~Narasimham, N.~Ito, S.~Kawakami and S.~Miyake,
  Pramana {\bf 5}, 59 (1975).
  doi:10.1007/BF02846033

\bibitem{proceeding}
M.R.Krishnaswamy et al, Proceedings XXIII International Conference on High  Energy Physics, Berkeley (ed.) S Loken (World Scientific, 1986).
 

\bibitem{Murthy:2013uca} 
  M.~V.~N.~Murthy and G.~Rajasekaran,
  Pramana {\bf 82}, 609 (2014)
  Erratum: [Pramana {\bf 88}, no. 4, 60 (2017)]
  doi:10.1007/s12043-014-0718-5, 10.1007/s12043-017-1365-4
  [arXiv:1305.2715 [hep-ph]].

\bibitem{Baring:2015sza} 
  M.~G.~Baring, T.~Ghosh, F.~S.~Queiroz and K.~Sinha,
  Phys.\ Rev.\ D {\bf 93}, no. 10, 103009 (2016)
  doi:10.1103/PhysRevD.93.103009
  [arXiv:1510.00389 [hep-ph]].

\bibitem{Boddy:2018qur} 
K.~Boddy, J.~Kumar, D.~Marfatia and P.~Sandick,
Phys.\ Rev.\ D {\bf 97}, no. 9, 095031 (2018)
doi:10.1103/PhysRevD.97.095031
[arXiv:1802.03826 [hep-ph]].


\bibitem{Evans:2016xwx} 
N.~W.~Evans, J.~L.~Sanders and A.~Geringer-Sameth,
Phys.\ Rev.\ D {\bf 93}, no. 10, 103512 (2016)
doi:10.1103/PhysRevD.93.103512
[arXiv:1604.05599 [astro-ph.GA]].


 \bibitem{Huniverse} Y. Mambrini, {\it Histories of Dark Matter in the
 Universe},
 www.ymambrini.com/My$\stackrel{~}{_-}$World/Physics$\stackrel{~}{_-}$files/Universe.pdf .
 



\end{thebibliography}
\end{document}